# NEXTFORMER: A CONVNEXT AUGMENTED CONFORMER FOR END-TO-END SPEECH RECOGNITION

*Yongjun Jiang, Jian Yu, Wenwen Yang, Bihong Zhang, Yanfeng Wang*

AI Interaction Division, Tencent PCG
{yjunjjiang, swordyu}@tencent.com

## ABSTRACT

Conformer models have achieved state-of-the-art(SOTA) results in end-to-end speech recognition. However Conformer mainly focuses on temporal modeling while pays less attention on time-frequency property of speech feature. In this paper we augment Conformer with ConvNeXt and propose Nextformer structure. We use stacks of ConvNeXt block to replace the commonly used subsampling module in Conformer for utilizing the information contained in time-frequency speech feature. Besides, we insert an additional downsampling module in middle of Conformer layers to make our model more efficient and accurate. We conduct experiments on two opening datasets, AISHELL-1 and WenetSpeech. On AISHELL-1, compared to Conformer baselines, Nextformer obtains 7.3% and 6.3% relative CER reduction in non-streaming and streaming mode respectively, and on a much larger WenetSpeech dataset, Nextformer gives 5.0%~6.5% and 7.5%~14.6% relative CER reduction in non-streaming and streaming mode, while keep the computational cost FLOPs comparable to Conformer. To the best of our knowledge, the proposed Nextformer model achieves SOTA results on AISHELL-1(CER 4.06%) and WenetSpeech(CER 7.56%/11.29%).

## 1. INTRODUCTION

In automatic speech recognition area, end-to-end(E2E) based systems have shown large improvements and achieved state-of-the-art results[1, 2] in recent years. Three modeling paradigms are applied in E2E systems: connectionist temporal classification(CTC)[3], neural transducer[4, 5] and attention based encoder-decoder (AED)[6, 7, 8]. Currently, CTC/AED models which combine the advantage of both system have shown promising result in ASR[9, 10, 11, 12].

For popular AED and neural transducer system, different kinds of model structure can be used. Such as long short term memory(LSTM)[7] or convolutional neural networks (CNNs)[13]. CNNs can capture local and delicate information with the drawback of needing many layers to capture global context. Transformer[14], first proposed in NLP area, was successfully transfered to E2E ASR[8] since it can model global context which is suitable for speech recognition. Recently, a novel structure Conformer[15] which combined CNN and self-attention was proposed. With this design, Conformer can capture global context and local correlation at the same time. Compared to Transformer, Conformer can get further improvement of modeling capability, became a standard choice for E2E ASR[1, 2, 12].

We focus on AED system since it was more commonly used. For Conformer based AED, an encoder and a decoder was applied. The encoder consists of a subsampling module which contains two Conv2d with stride to subsample audio frames, and multiple Conformer layers to extract fine-grained temporal feature as encoder output. The decoder consists of multiple Transformer layers to do auto-regressive decoding and inherent language modeling.

Despite the success of Conformer based AED, when inspect Conformer encoder, one can find that most of model capability was assigned to temporal modeling which is important for speech because speech is time series signal. But different with text, speech feature is time-frequency so frequency domain also plays another indispensable role. In Conformer encoder, the subsampling module directly manipulates time-frequency audio feature, then converts it to temporal feature for subsequent Conformer layers. As a consequence, only use two convolutions in subsampling module is a lack of capacity for time-frequency feature.

To pay more attention to time-frequency property of speech feature and make model more efficient and accurate, we propose Nextformer in this paper. Our main contributions are: first we design stacks of ConvNeXt[19] block to replace the role of subsampling module which is shown to be effective, second to relief the additional computational cost bring by ConvNeXt and improve modeling accuracy, we insert an additional downsampling module in the middle of Conformer layers. Experimental results show that our proposed Nextformer can get lower CERs than Conformer on AISHELL-1 and WenetSpeech both in non-streaming and streaming mode, while keep computational cost FLOPs comparable to Conformer.

The rest of this paper is organized as follows. In Section2 we review previous work. Section3 introduces detail of Nextformer. In Section4 we present experiment results. And finally we conclude this paper in Section5.

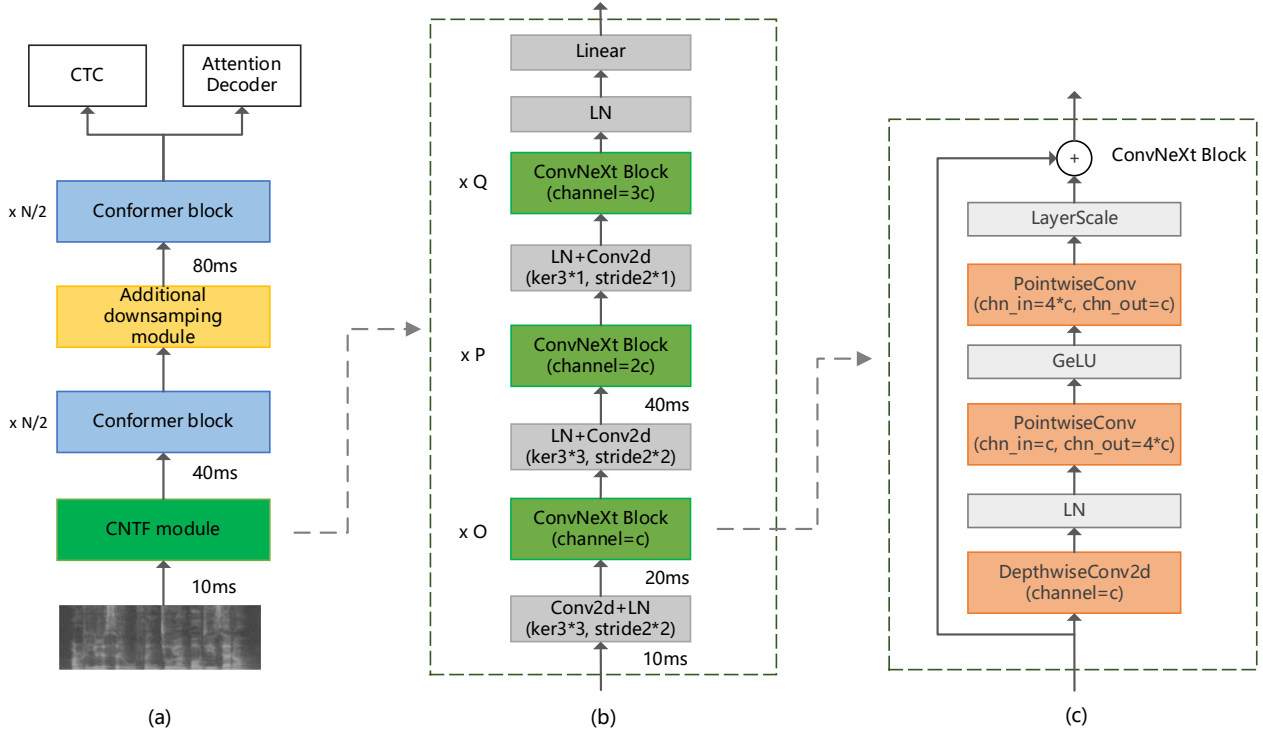

**Fig. 1**: *(a) The Nextformer based CTC/AED architecture. (b) ConvNeXt time-frequency(CNTF) module, this module applys time stride twice, so output frame rate is 40ms. (c) ConvNeXt block, which input and output channel is c.*

## 2. PREVIOUS WORK

Some methods have been proposed to address the time-frequency property of speech. In [9] a VGG net architecture which compose of 4-layer CNN and 2-layer MaxPool was used. Although deep CNNs, it's not very much different with subsampling module. In [16] a MobileNetV2-like nets was added to spectrogram. But their work focuses on multi-stream and multi-scale feature extraction. In deep stream they used many strides, which may heavily blur spectrogram.

The most related work to additional subsampling is [17,18]. In [17] they used progressive downsampling which contains in Conformer block and activate in the first two conformer layers to perform subsampling. In [18] they used a temporal U-Net architecture to first downsample to 80ms and then recover to 40ms. In this paper we use a simple yet effective additional downsampling module.

## 3. NEXTFORMER

The proposed Nextformer is based on CTC/AED system, we use Nextformer encoder to replace Conformer encoder while keep decoder as the same. Note one can easily transfer Nextformer encoder to neural transducer or other E2E systems.

### 3.1. System overview

The pipeline of our system shows in Figure 1(a). Assume we have speech feature $X = [x_1, x_2, ..., x_T]$ with frame $T$ and corresponding target label $Y = [y_1, y_2, ..., y_L]$ with length $L$. Processing $X$ with Nextformer encoder we get hidden encoder representation $H = [h_1, h_2, ..., h_M]$. Where Nextformer encoder consists of one ConvNeXt time-frequency module, two N/2 Conformer blocks and one additional downsampling module. For joint CTC/AED training, the objective is:

$$L = \alpha L_{ctc}(H, Y) + (1 - \alpha) L_{att}(H, Y) \quad (1)$$

where $L_{ctc}(H,Y)$ is CTC loss, $L_{att}(H,Y)$ is AED loss and $\alpha$ is a hyper-parameter to balance the weight of different loss.

### 3.2. ConvNeXt time-frequency(CNTF) module

ConvNeXt[19], a ResNet-like CNN structure, borrows a lot of successful ideas from Transformer, is shown to be competitive with Swin Transformer[20] in many computer vision tasks[19]. Compared to ResNet[21], ConvNeXt alters in other aspects such as larger kernel size and depthwise convolution, make it accurate and efficient. We use ConvNeXt here and design a module for delicately processing time-frequency speech feature and to replace the subsampling module in Conformer based AED.

First we introduce the structure of ConvNeXt block. As show in Figure 1(c), a ConvNeXt block with input channel and output channel $c$, which composes of a depthwise convolution, two pointwise convolutions, a LN layer, a GeLU activation layer and a LayerScale[22]. Here depthwise convolution is two dimensional as speech feature has time-frequency dimensions just like image, and we use a large kernel size(7*7 in this paper) as recommended by [19]. With the inverted bottleneck structure, preposing depthwise convolution makes computation efficient. The two pointwise convolutions act as the role similar to FeedForward sub-block in Transformer. For more details please refer to [19].

The CNTF module is stacks of ConvNeXt block with downsampling layers, as shown in Figure 1(b). Between different downsampling layers, we combine several ConvNeXt blocks as a stage, the block num per stage is $O, P, Q$. We design three stages in total, and progressively increase the channel size of each stage as $c, 2c, 3c$. Note that with this design and downsampling process, we can keep the FLOPs of each stage roughly the same which can give better results in previous experiment. The input is time-frequency speech feature with framerate 10ms. We first downsample it both in time and frequency axis to reduce computational complexity for afterward steps, this is done by a Conv2d with stride 2*2, then a layer norm is applied for stabilize training. The second downsampling layer is almost the same with the difference of changing position between convolution and layer norm. The third downsampling layer only downsample on frequency axis since we find that downsampling time here will harm results so we leave time downsamling in the additional downsampling module. The framerate here is 40ms. To avoid overfitting, we add stochastic depth[19] to each ConvNeXt block. Finally we use a linear layer to convert channel-time-frequency output to time-channel output for subsequent Conformers.

We also attempt to stack deeper CNNs to see whether deeper CNNs can act as the same role of CNTF but this gives worse results, which can be found in our experiment.

### 3.3. Additional downsampling module

We use additional downsampling module to reduce the overall computational cost of Nextformer. We insert this module in middle of Conformer blocks as this can balance accuracy and cost. For additional downsampling module, we use a FSMN-memory-like[23] downsampling layer, a Swish activation layer and a LN layer. The FSMN-memory-like downsampling layer is a lightweight layer with weight num 2 and stride 2, formulizes as:

$$h_{t'}^l = \sum_{i=0}^{1} w_i h_{t-i}^{l-1}, t\%2 = 1 \quad (2)$$

where $w$ is weight and $h^{l-1}$ represents inputs. This downsampling layer is slightly better than average pooling with negligible parameters compared to Conv1d.

## 4. EXPERIMENTS

### 4.1. Datasets

To evaluate our model, we conduct experiments on two open-source mandarin datasets, namely AISHELL-1[24] and WenetSpeech[25]. AISHELL-1 is a close-talk recorded speech corpus consists of a 150 hours training set, a 10 hours development set and a 5 hours test set. WenetSpeech is a large scale multi-domain mandarin corpus which contains a 10005 hours of high-quality labeled speech, a 20 hours Dev set, a 23 hours Test_Net set and a 15 hours Test_Meeting set.

### 4.2. Training setups

For AISHELL-1 we use the speed perturbation with factors of 0.9, 1.0 and 1.1 to augment training data. For WenetSpeech we use the 10005h training set only. For all experiments, we use 80-dimension log-Mel filterbank features, which are computed with a 25ms window and shifted every 10ms. A global mean and variance normalization is applied. We use SpecAugment on AISHELL-1 and WenetSpeech.

*Table 1: Model configurations for Conformer*

| Model | enc/dec | dim/head | params | FLOPs |
|---|---|---|---|---|
| Conformer-S | 12/6 | 256/4 | 46.3M | 11.0G |
| Conformer-L | 12/6 | 512/8 | 116.9M | 30.9G |

*Table 2: Model configurations for Nextformer, other parameters are the same as Table 1 listed.*

| Model | channel | O/P/Q | params | FLOPs |
|---|---|---|---|---|
| Nextformer-S | 56 | 3/3/3 | 46.1M | 10.9G |
| Nextformer-L | 104 | 3/3/3 | 115.1M | 30.9G |

For baseline models, we use Conformer-S and Conformer-L, as shown in Tabel 1. For the subsampling module in Conformer models, we use two Conv2d with kernel size 3*3 and stride 2*2, result in 40ms framerate. The *ffd_dim* is equal to *4*dim* and all kernel size in Conformer block is 15. We use a 10s long speech to calculate the FLOPs, and only consider encoder for simplicity. For Nextformer, as shown in Table 2, we train Nextformer-S and Nextformer-L.

To validate the generalization performance of our model, we apply both non-streaming and streaming mode. For streaming mode, we use the similar method in [26] which consist of dynamic chunk training and causal convolution in Conformer blocks, and we make the *chunksize* to be even to support additional suabsampling. In Nextformer, we make all convolutions in CNTF causal for not leaking future context.

For ALSHELL-1 we apply Adam optimizer and warmup learning rate scheduler with 25,000 warmup steps and peak learning rate 5e-4, we train 100 epochs for non-streaming model and 120 epochs for streaming model. For WenetSpeech we apply Adam optimizer and propose a new learning rate scheduler *WCE-lr*, that is, we first warmup learning rate to peak, then keep it constant, at last exponentially decay it per epoch. As for *WCE-lr* in WenetSpeech, we set 25,000 warmup steps, peak learning rate 5e-4, exponential decay ratio 0.6 and begin exponential decay in epoch 16. We train 25 epochs for WenetSpeech both in non-streaming and streaming mode.

Table 3: CERs results on AISHELL-1.

| Mode | Method | Model | dev | test |
|---|---|---|---|---|
| non-streaming | beam search | Conformer-S | 4.39 | 4.89 |
| | | Nextformer-S | 4.20 | 4.55 |
| | WNARS | Conformer-S | 4.09 | 4.38 |
| | | Nextformer-S | **3.86** | **4.06** |
| streaming | beam search | Conformer-S | 4.97 | 5.50 |
| | | Nextformer-S | 4.72 | 5.25 |
| | WNARS | Conformer-S | 4.65 | 5.04 |
| | | Nextformer-S | **4.36** | **4.72** |

We set attention dropout to 0.1 and label smooth to 0.1. And in Nextformer we set stochastic depth to 0.1. CTC weight $\alpha$ is set to 0.3. All models are trained on Espnet[1] with 8 NVIDIA's Tesla A100 GPUs.

### 4.3. Result on AISHELL-1

We train Conformer-S and Nextformer-S on AISHELL-1. Then evaluate them with two decoding methods: joint CTC-attention beam search and WNARS[12], a wfst-CTC decoding method with attention rescoring. For WNARS, we train a 3-gram LM based on the corpus in training set. Results are shown in Table 3. As we can see, in non-streaming mode, comparing to Conformer-S, Nextformer-S obtains 7% and 7.3% relative CER reduction on testset with joint CTC-attention beam search and WNARS respectively. In streaming mode, the relative CER reduction is 4.5% and 6.3%, which we use a 640ms decoding chunk length. Using Nextformer-S, we can reduce the CER on AISHELL-1 test set to 4.06%, which is the SOTA result at present. Note that WNARS has obvious improvement compared to joint CTC-attention beam search, we'll use WNARS decoding method in subsequent sections.

### 4.4. Results on WenetSpeech

We move to a 10000h WenetSpeech dataset which is larger and more complicated. Here we use Conformer-L and Nextformer-L configurations. For decoding, we use training set corpus to build a 4-gram LM and split text with jieba[2]. Results are shown in Table 4. In non-streaming mode, Nextformer-L has 5.0% relative CER reduction and 6.5% relative CER reduction compared to Conformer-L baseline. In streaming mode, Nextformer-L gets 7.5% relative CER reduction and 14.6% relative CER reduction compared to Conformer-L, bigger gains compared to non-streaming mode. We guess that in streaming mode, Nextformer can get better usage of limited left time-frequency context. Besides, we reduce CER on Test_Net to 7.56% and Test_Meeting to 11.29%, which is the SOTA results on WenetSpeech to best of our knowledge.

---

[1] https://github.com/espnet/espnet
[2] https://github.com/fxsjy/jieba

Table 4: CERs results on WenetSpeech

| Mode | Model | Dev | Test_Net | Test_Meeting |
|---|---|---|---|---|
| non-streaming | Conformer-L | 7.05 | 7.95 | 12.08 |
| | Nextformer-L | **6.95** | **7.56** | **11.29** |
| streaming | Conformer-L | 7.73 | 9.56 | 15.06 |
| | Nextformer-L | **7.44** | **8.84** | **12.86** |

### 4.5. Component study

We make several component studies of Nextformer on AISHELL-1. As listed in Tabel 5, first we validate the effect of CNTF module and additional downsampling module, we can see both of the two modules are beneficial to final result. Then we replace CNTF module with 8-layer CNNs(details are shown in Tabel 6) to further verify the modeling ability of CNTF, we add residual connection between the second and fourth CNNs block and the FLOPs is 48.8G of encoder. From Table 5, we can see CNTF is better than 8-layer CNNs with a quarter of FLOPs. Third we study the result of different position of additional downsampling module, we find that putting additional downsampling module after middle of Conformer layers gives the best result.

Table 5: Component study results on AISHELL-1

| Models | CER |
|---|---|
| Nextformer-S | 4.06 |
| - CNTF module | 4.18 |
| - additional downsampling module | 4.14 |
| Nextformer-S | 4.06 |
| replace CNTF with 8-layer CNNs | 4.19 |
| Nextformer-S | 4.06 |
| additional subsamp. on Cfm. layer 5 | 4.13 |
| additional subsamp. on Cfm. layer 7 | 4.10 |

Table 6: Detail of 8-layer CNNs to replace CNTF module

| Operator | Channel | Stride | Repeat |
|---|---|---|---|
| Conv2d+ReLU | 256 | 2 | 1 |
| Conv2d+ReLU | 256 | 1 | 3 |
| Conv2d+ReLU | 256 | 2 | 1 |
| Conv2d+ReLU | 256 | 1 | 3 |

## 5. CONCLUSIONS

We propose Nextformer in this paper. We design ConvNeXt time-frequency module to extract delicate information in speech feature. We apply additional downsampling module to reduce the overall FLOPs and further improve accuracy. Experimental results show the effectiveness of our model. We will apply our model to more datasets and more ASR systems such as neural transducer and hybrid system in the future.